\documentclass{article}
\begin{document}
\title{How the BAL quasars are quiet.}
\author{Jacques Moret-Bailly 
\footnote{Laboratoire de physique, Université de Bourgogne, BP 47870, F-21078 
Dijon cedex, France.
Email : Jacques.Moret-Bailly@u-bourgogne.fr
}}
\maketitle 

\medskip
Pacs 42.65.Dr Stimulated Raman scattering, 98.54.Aj Quasars 

\medskip

 \abstract{Taking the coherent light-matter interactions into account, all properties of the quasars 
known by the author (quietness of the BAL quasars, Lyman $\alpha$ forest, iron in the high-z 
quasars...) are explained supposing that the kernel of the quasar is surrounded by a halo containing 
an accretion disk and magnetic satellites (no dark matter or fast jet, no variation of the fine structure 
constant, no strange source of iron...). }

\section{Introduction}
Recent observations show two discrepancies in the standard interpretation of the spectra of quasars: 

i) the relative Doppler or expansion frequency shift of spectral lines is strictly constant, while the 
observations show slight variations \cite{Webb}; the explanation by a variation of the fine structure 
constant is hardly credible;

ii) the concentration of iron in quasars whose spectrum is largely shifted requires a generation of 
this metal in old stars \cite{Hasinger}, implying a too large increase of the age of the 
Universe.

The standard model seems unable to explain that the loud quasars do not have broad lines.

Twenty years ago, it seemed difficult to interpret various observed redshifts by the Doppler effect 
\cite{Reboul}, but the lack of a credible alternative led to find solutions which do not convince all 
astrophysicists.

\medskip
In 1968, Giordmaine et al. \cite{Giordmaine} observed frequency shifts in experiments using high 
power ultrashort laser pulses; this observation is now common, used for analysis and called 
''Impulsive Stimulated Raman Scattering'' (ISRS)\cite{Weiner,Dougherty,Dhar}. The theory of ISRS 
shows that this effect has no intensity threshold \cite{Yan}. The intensities of the lasers make ISRS 
non-linear, so that the frequency shifts depend on these intensities; at low intensities, the effect 
becomes linear, so that it is similar to a Doppler effect, except for a small dispersion. This qualitative 
difference justifies a new name: "Coherent Raman Effect on time-Incoherent Light" (CREIL).

In the next section, the properties of the CREIL are recalled. In the following sections, the influence 
of each of these properties on the standard model of quasars is given, solving very simply some 
problems, in particular the noisy quasars cannot show broad spectral lines.

\section{Properties of the  "Coherent Raman Effect on Incoherent Light" (CREIL).}
Any alternative to the Doppler effect in the vacuum requires a time-incoherence of the light: If, during 
a certain time, the received frequency of a coherent light is lower than the emitted frequency, the 
number of cycles, that is the number of wavelengths between the source and the receiver is 
increased; as the wavelength is an unit of length, the distance is increased, it is a Doppler effect. 
Thus, we understand that the length of the pulses of light is critical, the pulses of light must verify the 
condition set by G. L. Lamb when the length of the pulses is a condition for the observation of an 
optical effect: the "ultrashort light pulses", must be "shorter than all relevant time constants" 
\cite{GLamb}. 

In an homogeneous gas, two relevant time constants must be longer than the pulses which make the 
usual incoherent light, that is longer than about 10 nanoseconds:

i) the collisional time $\tau$ must be longer than 10 ns, for instance $\tau=100ns$.
For hydrogen atoms at 10000K, the density of atoms $N$ is $N\approx 1,7.10^{21}$ 
atomes/m$^3$; for H$_2$ at 100K, $N\approx 2,4.10^{22}$; these values are orders of 
magnitude because the mean diameter of the atoms depends on the criterion used to define a 
collision. This condition allows that all molecules on a wave surface remain in phase and scatter 
coherent waves which rebuild wave surfaces identical to the incident wave surfaces: no blur of the 
images.

ii) The period corresponding to the Raman transition must be larger than 10 ns, that is in the 
radiofrequencies range. As the period of the beats of the incident and scattered beams is 
much longer than the duration of the pulses, the beats do not appear, the incident and 
scattered beams interfere into a single frequency beam; the frequency shift is proportional to 
the ratio of the amplitudes and to the Raman frequency.

Although the coherent scattering is much more powerful than an incoherent Raman scattering, the 
previous conditions make CREIL much weaker than ISRS which uses dense matter and infrared 
Raman frequencies; an other source of weakness is that the populations of the levels involved in the 
Raman transition are nearly the same because they are close, so that the opposed virtual transitions 
nearly cancel the shifts that they produce. Thus, it seems impossible to get in the labs long enough 
paths to observe the CREIL while the Universes provides long paths in low pressure gases.

Previous papers \cite{Moret1,Moret2,Moret3,Moret4} show that the cosmological redshift may be 
produced by CREIL if there are, in the average, some active molecules per litre in the intergalactic 
space; it is a very rough evaluation because a precise computation would require for each molecule 
present in the space, the knowledge of a lot of traces of Raman polarisation tensors.

Suppose that a redshifting gas has absorption lines; the absorption by a line occurs from the 
redshifted frequency to the absolute frequency; if the redshift is large, the line is wide and weak. If 
the gas has many absorption lines, the lines are mixed, they cannot be observed.

If ISRS and CREIL were simple Raman effects, the active molecules would be excited, and, without 
collisions, their de-excitations would be slow; ISRS and CREIL are parametric effects which mix the 
described Raman effect with an other which de-excite the molecules and blue-shift (an) other 
beam(s); in ISRS, the other beam is a laser beam, in CREIL, its role may be plaid by the thermal 
radiation which is amplified. In this last case, all frequencies are low, resonant.

\section{Distance (age) of the quasars}
The distance and the age of the quasars is deduced from the redshift of the sharp emission or 
absorption lines, which are the most redshifted lines. It was remarked that the neighbourhood of the 
most redshifted quasars contains a lot of matter, in particular molecular hydrogen detected by the 21 
cm nuclear spin coupling line. It is reasonable to think that where the pressure is low, the molecular 
hydrogen is partly ionised by the UV into H$_2^+$. This gas cannot be detected by its absorption 
or emission lines because it has many relatively weak lines which are shifted during their absorption 
or emission by the CREIL; if the pressure is high enough to produce collisions which forbid the 
CREIL, these collisions destroy H$_2^+$ too quickly to allow its observation.

The CREIL distorts the standard scale of distances deduced from the redshifts. This scale remains 
probably good in the intergalactic space, because there, the redshift is produced by expansion or by 
the CREIL in a nearly homogenous medium; but, where there is much molecular hydrogen, in 
particular in the neighbourhood of the quasars the distances must be strongly reduced.

\section{Absence of broad lines in the loud quasars}
Suppose that a quasar has an accretion disk larger, but similar to Saturn's disk, except maybe for an 
inner part which protects the remainder from burning. The friction of the stones on the halo of the 
quasar charges them, just as the drops of rain are charged in our clouds. While the lightnings 
discharge a volume of our clouds, the discharges of regions of the disk are limited to its plane, so that 
the electromagnetic field radiated by sheet-shaped currents propagate more in the direction of the 
axis of the disk.

\medskip
The disk is only partly and locally discharged by the lightnings; as the disk is inhomogeneous, the sign 
of its charges depend on the distance to the quasar; thus, the rotation of the disk produces circular 
currents which induce a magnetic field similar to the field introduced by de Kool and 
Begelman\cite{Kool}: near the disk, its orientation may change, so that it may have zero values, in 
particular if the disk has gaps.

Raman transitions inside a Zeeman structure of the hydrogen atom require a principal quantum 
number larger than one, that is the atom must have been excited by a Lyman absorption; then the 
excited atom is active for CREIL were the magnetic field is not null; for atoms pumped by a Lyman 
$\alpha$ transition, neglecting the electronic spin, and supposing that the de-exciting Raman 
component is very strong, the computation of the CREIL requires only the trace of a single, simple 
Raman polarisability tensor. If the Lyman lines are redshifted while they are absorbed, they appear 
so wide and weak that they cannot be  seen. 

If a line of sight makes large enough an angle with the axis of the quasar, it crosses places where the 
magnetic field is nearly zero, so that, as there is no CREIL in these places, the absorption of the 
Lyman lines is stable in the spectrum and the lines appear. The large width of the lines may be a 
consequence of a low CREIL (by a low magnetic field, or collisional decrease), or, as this process is 
close to the quasar, it may depend slightly on the line of sight. This description of the origin of the 
BAL applies to the BEL.

\medskip
Thus, the loudness, the presence of broad lines is not a property of the quasars, but depends on the 
direction of observation: BAL quasars, observed in the direction nearly perpendicular to the axis of 
the disk, cannot radiate much electromagnetic field in this direction so that they are ''quiet''.

\section{The Lyman forest }
In the standard model, the Lyman forest corresponds to clouds, that is to variations of the density of 
hydrogen atoms. It may be the result of the absorption in a chemically homogenous medium and a 
space-variable redshifting power of the gas resulting of variations of the magnetic (or electric) field. 
But how could a space-variable magnetic field be generated in the far halo?
Maybe by electric instabilities in the plasma, or it could be the magnetic part of an extremely low 
frequency electromagnetic field radiated by the quasar and propagating very slowly in the plasma, 
more probably by magnetised satellites of the quasar.

In one of these hypotheses, the modulation of the absorption which structures the Lyman forest is 
self-amplifying: 
 Follow the light; if, in a region, the magnetic field is very low, there is no redshift, the Lyman 
$\alpha$ line saturates, so that, in the following region there is no Lyman $\alpha$ pumping, thus no 
CREIL even if a magnetic field appears; a high magnetic field is necessary to shift the spectrum by a 
small number of remaining CREIL-active atoms ( hydrogen pumped by other Lyman lines, and other 
atoms). When the line is shifted of a fraction of its linewidth, the Lyman $\alpha$, the CREIL and the 
redshift become strong, the absorption is spread and seems low in the spectrum; a return to the 
saturation requires a low magnetic field. This discussion is valid where the forest is saturated, but its 
results remain true elsewhere: only small fluctuations of the field may induce the forest.

\medskip
Two effects explain that the density of lines in the high redshift region of the forest generally 
decreases: i) the density of hydrogen being higher and its illumination higher, the CREIL is larger, the 
spectrum is spread, so that the decrease is an artefact; ii) in this region, some small satellites may fall 
into the accretion disk.

Along a line of sight, the frequencies are shifted near the magnetic satellites while the Lyman lines are 
written far from these satellites. When an eclipse is observed by a variation of the intensity of the 
quasar, if small variations of the redshift can be detected, the gap corresponding to the satellite in the 
Lyman forest may be found. It could help a study of the satellite and the quasar.

\section{High redshift quasars}
The high redshift quasars appear in regions of the Universe in which clouds of molecular hydrogen 
are detected by the nuclear spin coupling transition at 21 cm. They seem dusty too.

If the pressure of molecular hydrogen allows a very long collisional time, the UV ionises a part of the 
hydrogen into H$_2^+$ which is active in CREIL, thus cannot be observed from its absorption or 
emission spectra. The CREIL increases the redshift of the quasar; it amplifies the thermal radiation, 
possibly up to 100K, so that this thermal radiation may be confused with radiation of hot dust.

\medskip
If a galaxy appears next to a quasar, and has a lower redshift, it may, however, be further than the 
quasar. Thus its line of sight may cross the neighbourhood of the quasar, so that absorptions of metal 
lines may have the same origin, appear with the same redshift.

\section{Conclusion}
This draft intends only to show that taking the Coherent Raman Effect on time-Incoherent Light 
(CREIL) into account solves very easily, without any new physics or new matter, problems which 
appear in the study of the quasars:
- the absence of BAL lines in the loud quasars is deduced from the hypothesis of a disk similar to 
Saturn's;
- the discrepancies in the observed spectra may be produced by the small dispersion of the 
CREIL;
- the quasars are younger than they seem, so that they can contain metals made in old stars.

The use of CREIL allows to find several explanations of a particular observation; the present 
qualitative study must be specified quantitatively to choose the best explanation, maybe the standard 
explanation: for instance the thermal radiation may be partly produced by hot dust, partly by CREIL. 
Having several possible solutions is better than having none !

The model of quasars which takes CREIL into account uses only commonly observed types of 
astrophysical objects and simple regular physics; with much less work, it seems to give 
interpretations of observations on the quasars better than the standard model; it does not require 
dark matter, but if this matter is necessary for the gravitational stability, it proposes partially ionised 
H$_2$. Logically, is it acceptable to neglect CREIL {\it a priori}?

The present demonstration of the necessity to take CREIL into account in astrophysics weakens the 
two main proofs of the big bang, expansion and 2.7K radiation.

\end{document}